Note: For the final published version of the paper please refer to: Baldwin J., Dehghantanha A. (2018) Leveraging Support Vector Machine for Opcode Density Based Detection of Crypto-Ransomware. In: Dehghantanha A., Conti M., Dargahi T. (eds) Cyber Threat Intelligence. Advances in Information Security, vol 70. Springer, Cham, DOI https://doi.org/10.1007/978-3-319-73951-9_6

# Leveraging Support Vector Machine for Opcode Density Based Detection of Crypto-Ransomware

James Baldwin, Ali Dehghantanha

Department of Computer Science, School of Computing, Science and Engineering, University of Salford, UK

**Abstract**

Ransomware is a significant global threat, with easy deployment due to the prevalent ransomware-as-a-service model. Machine learning algorithms incorporating the use of opcode characteristics and Support Vector Machine have been demonstrated to be a successful method for general malware detection. This research focuses on crypto-ransomware and uses static analysis of malicious and benign Portable Executable files to extract 443 opcodes across all samples, representing them as density histograms within the dataset. Using the SMO classifier and PUK kernel in the WEKA machine learning toolset it demonstrates that this methodology can achieve 100% precision when differentiating between ransomware and goodware, and 96.5% when differentiating between 5 crypto-ransomware families and goodware. Moreover, 8 different attribute selection methods are evaluated to achieve significant feature reduction. Using the CorrelationAttributeEval method close to 100% precision can be maintained with a feature reduction of 59.5%. The CFSSubset filter achieves the highest feature reduction of 97.7% however with a slightly lower precision at 94.2%.

Using a ranking method applied across the attribute selection evaluators, the opcodes with the highest predictive importance have been identified as FDIVP, AND, SETLE, XCHG, SETNBE, SETNLE, JB, FILD, JLE, POP, CALL, FSUB, FMUL, MUL, SETBE, FISTP, FSUBRP, INC, FIDIV, FSTSW, JA.  The MOV and PUSH Opcodes, represented in the dataset with significantly higher density, do not actually have high predictive importance, whereas some rarer opcodes such as SETBE and FIDIV do.

# 1. Introduction

In their December 2016 quarterly threat report [1] McAfee referred to 2016 as the "year of ransomware; the FBI estimated that $1Billion of ransom demands were paid in 2016 representing a 400% increase from the previous year, and the cost of the average ransom demand doubled [2]. The rise of the ransomware-as-a-service (RaaS) model provided cybercriminals with the ability to distribute ransomware with very little technical knowledge [3] in addition to the potential for huge financial returns for both the distributors, and the developers within the model. (In July 2016 Cerber generated US $195,000 revenue for its distributors with a 40% cut of that figure going to the developer) [4].

Europol, in 2016, reported that ransomware had become the primary concern for European law enforcement agencies with Cryptoware the most prominent malware threat [5]. Again, in 2017, they concluded that ransomware continued to be "one of the most prominent malware threats in terms of the variety and range of its victims and the damage done" [6]. In the recently commissioned 2017 Ransomware Report, 88% of survey respondents who has been affected by ransomware in the previous year had encountered crypto-ransomware [7].

In 2017 ransomware achieved global news coverage due to the WannaCry [8] and subsequent Petya [9] outbreaks. Due to the nature of the WannaCry outbreak many high profile global organisations such as the UK National Health Service, Spanish telecommunications company Telefónica, and logistics company Fed-Ex. were subject to severe disruption [6]. The scale of the infection and subsequent media coverage provoked discussion and reaction down from government level through to security vendors, enterprises and domestic audiences.

Current AV vendors that rely on static detection methods are struggling to contain the threat of ransomware due to the daily deployment of new variants, iterations and families. Recent new commercial software products have adopted heuristic detection methods for cyber defence purposes. For example Cybereason use behavioural techniques to protect consumer networks [10]; Darktrace employ advanced unsupervised machine learning for the protection of enterprise networks [11]; MWR have developed RansomFlare "as an effective countermeasure to the increasing threat of ransomware" [12]. Although malware threats and detection techniques are predominantly targeted towards Microsoft Windows systems, machine learning techniques are also applied to other platforms such as OS X [13], Android [14] and IOT (Internet of Things) [15].

Due to its recent significance and effect on the cyber threat landscape [16] this paper focuses on crypto-ransomware affecting Microsoft Windows systems only. It will use static opcode density analysis of crypto-ransomware and benign samples using Support Vector Machine (SVM) supervised machine learning techniques. It aims to achieve >95% precision when differentiating between ransomware and goodware, and when differentiating between 5 crypto-ransomware families and goodware. Moreover, 8 different attribute selection methods are evaluated to achieve feature reduction. This research will provide scope for further development, and practical or commercial application.

The rest of the paper is organised as follows: Section 2 reviews related works on the topic of machine learning, ransomware and malware detection; Section 3 describes the research methodology and the 4 phases that it comprises; Section 4 presents the results of the experiment; Section 5 presents the conclusion and discusses future works.

## 2. Related works and research literature

[17] demonstrated that the use of machine learning techniques could detect malware effectively and efficiently. They used dynamic analysis reports obtained from an online analysis service, Anubis, to evaluate different machine learning classifiers, with the best performing J48 Decision Tree classifier achieving precision of 97.3%, and a false positive rate (FPR) of 2.4%. In 2011 [18] proposed a framework for the automatic analysis of malware behaviour using machine learning. They embedded the observed behaviour in a vector space and applied clustering algorithms to achieve significant improvement over previous work in that area. [19] presented a comprehensive overview of the state-of-the-art analysis techniques and tools for use by a malware analyst. They concluded that both static and dynamic analysis tools were required to overcome the evasion techniques that malware authors were using. The following year, after a detailed analysis conducted on the current malware detection systems that included static, dynamic and hybrid malware techniques, [20] concluded that data mining and machine learning was also required to compliment the limitations of existing techniques to achieve better detection. [21] introduced a classification method that integrated static and dynamic to overcome the limitations associated with each technique. They demonstrated the importance of including both old and new malware samples to overcome the evasion techniques employed by malware authors. They achieved a minimum of 5% lower accuracy across all evaluated classifiers when using newer samples only. Similarly, in 2016 another comprehensive review was undertaken of techniques and tools for malware analysis and classification by [22]; it included processes for collection, static/dynamic analysis, feature extraction and machine learning classification.

Bilar [23] used static analysis techniques to statistically compare the distribution of opcodes within malware and goodware. He concluded that the more infrequent opcodes were a better indicator of malware compared to the more frequent opcodes. This statistical research has been cited frequently and is one of most important works relating to opcodes and their ability to be used for malware prediction. Bilar (2007) then extended his research to analyse opcode control flows which is now a widely used technique for malware detection. [25] presented a control flow-based method to extract opcode behaviour from executable files to improve on text-based extraction methods and obtain a more accurate representation of the executable file behaviour. The control flow method achieved a 2.3% improvement in accuracy over text-based methods using a KNN classification. [26] used a similar technique with additional feature filtering to achieve a 97% accuracy with a Random Forest classifier.
[27] proposed a malware classification system using approximate matching of control flowgraphs and a distance metric based on the distance between feature vectors of string-based signatures to achieve high accuracy.

Opcode density analysis has been extensively researched and applied to general malware detection. [28] used opcode density histograms to differentiate between malicious and benign PE files with 100% accuracy based on a 100 sample dataset (80 malicious /20 benign). [29] used dynamic analysis techniques to extract the opcodes from runtime traces and introduced SVM classifier filter techniques to perform dimension reduction.

Further dimension reduction/feature filtering research has included [30] who proposed a two-stage dimensionality reduction approach combining feature selection and extraction to substantially reduce the dimensionality of features for training and classification.

[31] used static analysis techniques for opcode n-grams and compared several feature selection methods and machine learning classifiers. They concluded that Principal Component Analysis (PCA) feature selection and Support Vector Machines (SVM) classification achieved the highest classification accuracy using a minimum number of features. [32] used feature reduction to identify the top-10 opcodes for malware detection and decreased the training time of a supervised learning algorithm by 91% with no loss of accuracy. [33] used feature reduction to identify 9 features that could distinguish malware from goodware achieving an accuracy of 99.60% using a Random Forest classifier.

There are several techniques that have been researched for the detection of ransomware as a specific family. EldeRan [34], was a ransomware classifier based on a sample's dynamic features; it achieved a True Positive Rate (TPR) of 96.3% with a low False Positive Rate (FPR) of 1.6%. UNVEIL [35] is another machine learning based system that uses a ransomware sample interacting with the underlying O.S. to achieve a True Positive Rate (TPR) of 96.3% and a zero False Positive Rate (FPR). Network related behaviour and Netflow data can also be used for ransomware detection as demonstrated by [36] and [37] in which they extracted conversation-based network traffic features to achieve a precision of 97.3% using the Decision Tree (J48) classifier. In 2016 [38] introduced 2entFox, a framework for ransomware detection based on 20 extracted filesystem and registry events. Using a Bayesian network model, it achieved an F-measure of 93.3%. Most recently in 2017, [39] used sequential pattern mining of filesystem, registry and DLL events to achieve 99% accuracy when differentiating between crypto-ransomware and goodware, and 96.5% when differentiating between the 3 different ransomware families. Using a novel approach to detect crypto-ransomware in IOT networks based on power consumption, [40] achieved a detection rate of 95.65%, and a precision of 89.19% when monitoring connected Android devices.

## 3. Methodology

The methodology can be divided into 4 key stages as summarised in Fig.1:

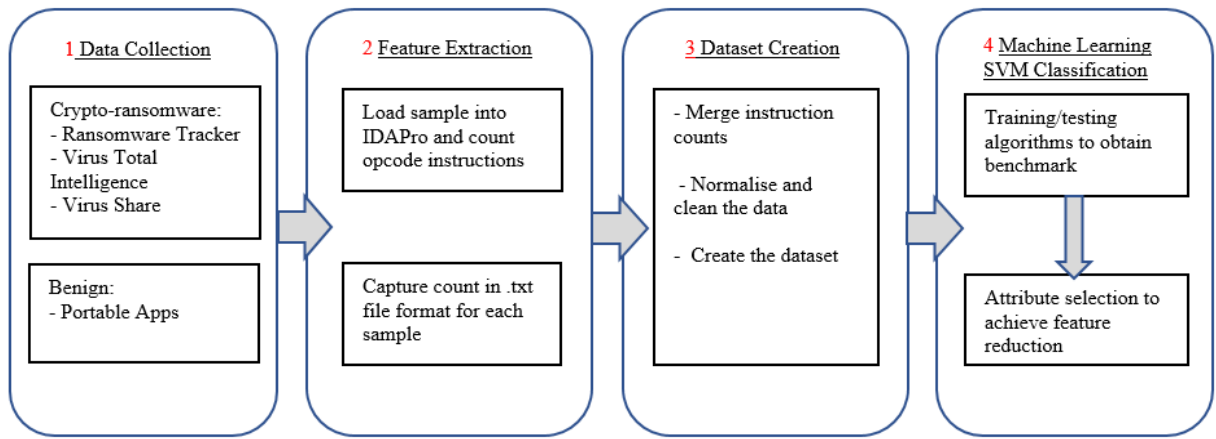

*Fig. 1 The 4 key stages of the research*

3.1    Data Collection

The ransomware families used for the experiment were identified from the Ransomware Tracker website [41], an online ransomware resource used by internet service providers (ISPs), computer emergency response teams (CERTs) and law enforcement agencies (LEAs). By studying the current tracked ransomware families focus was given to 5 crypto families only (Cerber, CryptoWall, Locky, TeslaCrypt, TorrentLocker) with the inclusion of historic samples where available to ensure that variant history and changes to the malicious code were included [21]. The Ransomware tracker website lists known Distribution, Payment and Command & Control hosts associated with each ransomware family; corresponding SHA256 sample references are documented. The SHA256 value for each identified sample was submitted to the Virus Total Intelligence platform [42] to facilitate quick, bulk downloading. The download list was reviewed to ensure that only the PE file format (Portable Executable) was downloaded, and other formats such as DLL (dynamic link library) were omitted. This was done to enable an accurate comparison with the goodware samples which were a similar PE format.

The benign (Goodware) executable samples for the dataset were obtained from the Portable Apps platform [43], a portable software solution that allows popular software to be installed and run from portable storage devices. The Portable Apps platform provides suitable counterpart samples in that the executable files are portable (or standalone), and ensure relevance across multiple OS versions [44]. Using a '*certutil*' command run against the 350 benign samples an output file was generated containing the SHA256 values for each sample. The SHA256 values were submitted in bulk to the VTI platform to determine the detection ratio for further analysis. Many of the samples were detected as infected by smaller antivirus vendors but the vendors that offered the highest protection e.g. BitDefender, F-Secure, Kaspersky , McAfee, Symantec, Trend Micro C [45] identified each of the goodware samples as clean, therefore the detection from the smaller vendors were deemed to be false positives, and discounted. Where a detection ratio of 4/total or higher was registered (regardless of vendor size) the samples were discarded.

3.2    Feature extraction

As the study focuses on the actual CPU instruction (opcode) and not the data that is to be processed (operand) the operand value has not been captured during the feature extraction process. Static analysis was performed using IDAPro with an InstructionCounter plugin to automate the opcode count process and export the results to text file format [46]. The InstructionCounter plugin was successfully installed and run correctly within a reduced-functionality, evaluation version of the latest IDAPro version 6.95 [47]; it was tested successfully in both a Windows 7 Professional SP1 and Windows 10 Home environments. On loading the sample into IDAPro the plugin was run and the output captured to a text file that was saved with a name corresponding to the SHA256 hash file for easy recognition. This process was repeated for all samples, both malicious and benign. Due to the limitations with the evaluation version of the latest IDAPro 6.95, only 32-bit PE files could be loaded – 64-bit benign samples files were discarded and not analysed.
Fig. 2 shows a sample feature extract using the InstructionCounter plugin.
The columns in the output file represent rank, count, density and, opcode. The total number of opcodes is also included which provide useful for reference and to enable recalculation of the density with more numerical precision within the dataset (see section 3.3.2).

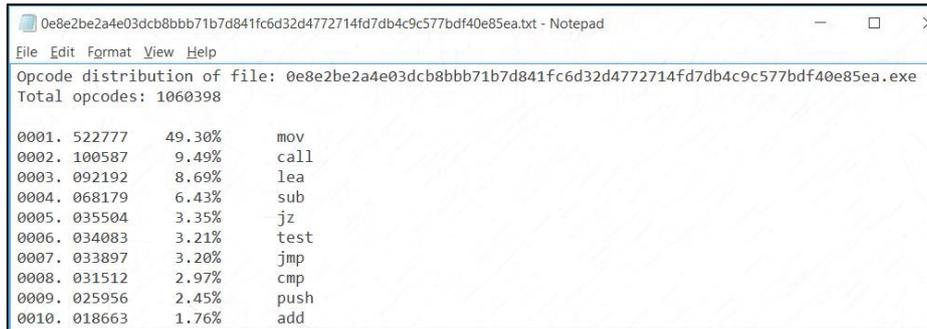

*Fig. 2 Sample InstructionCounter output file with top 10 opcodes, count and density*

### 3.3 Dataset creation

Each executable sample is represented in the dataset as a density histogram to reflect the percentage density of each opcode occurrence relative to the total number of opcode occurrences. This process removes any variance that is attributed to different application and code lengths. Consider the 2 samples in Table 1 which relate to the MOV opcode. The malicious and benign samples have a large variation in both the total number of opcodes, and extracted number of MOV opcodes. In both samples, however, MOV has a similar density and therefore are both equally ranked because of this.

*Table 1*
*Example of MOV Opcode density*

| Sample | MOV Opcode Rank | Total no. | Density | Total opcodes in sample |
|---|---|---|---|---|
| Malicious (Cerber) | 1 | 2,368 | 33% | 7,215 |
| Benign | 1 | 398,332 | 33% | 1,192,874 |

#### 3.3.1 Merging the data

A copy of all the malicious and benign output text files were placed in a single directory and merged into a single text file. The text file was opened in Microsoft Excel and using the 'Text to Columns' / 'Sort A to Z' / 'Remove Duplicates' features a master list of all the extracted opcodes (443) was obtained. This list was used to subsequently ensure that all records and features could be sorted alphabetically and therefore be correctly ordered.
Using a custom recorded VB Macro each output file was combined with the master opcode list, sorted, transposed and inserted into a master Microsoft Excel worksheet to create a raw representation of the data.

#### 3.3.2 Normalising the data
Additional processing in Excel was required to finalise the created dataset.

1. Each column was checked for correct alignment and sorting, samples were sorted by opcode count, records analysed and removed if containing duplicate data.

The cell values were recalculated to increase the decimal place value to 8 as the InstructionCounter plugin had only exported to 2 decimal places. The increase in numerical precision was required as 8 decimal places provided a differentiation between no occurrence (i.e. 0 count) and low occurrence (e.g. 5), which in samples of high opcode count were both represented by 0.00.

2. An additional "class" column was added where 'good' represented goodware and 'malware' represented malware to provide a response/detection for the machine learning prediction.

3. The dataset values were linearly scaled (0, 1). This was done for two reasons: to avoid attributes in greater numeric ranges dominating those in smaller numeric ranges; to avoid large attribute value calculation errors associated with some kernel functions. [48]

4. The final step was to calculate average density for each attribute and sort high to low. Fig. 3 shows an example of the final sorted dataset with the attributes highlighted. (The 'class' attribute is usually listed at the end of the dataset and is referenced in WEKA using the 'last' parameter. It is not shown in this example)

*Fig. 3 Dataset sample*

### 3.3.3 OpCode breakdown

Almost 71 million opcodes were extracted from the benign samples with an average of 308,038 codes per sample. This is in contrast to almost 3.5 million codes extracted from the malicious samples, with an average of 13,147 codes per sample. (Table 2)
Fig. 4 also illustrates the opcode distribution for both classes with the malware samples predominantly using < 50 different opcodes, in comparison to goodware using between 100-200. It can be clearly seen that the structure for the malware samples is much simpler than the goodware samples with a lower overall count for both the number of extracted codes, and number of different opcodes used to perform the instructions.

*Table 2*
*Extracted opcode statistics*

| Class | Total extracted opcodes | Average no extracted opcodes per sample | Largest no codes |
|---|---|---|---|
| Goodware | 70,848,904 | 308,039 | 320 |
| Malware | 3,418,229 | 13,147 | 163 |
| | | | **433 Total** |

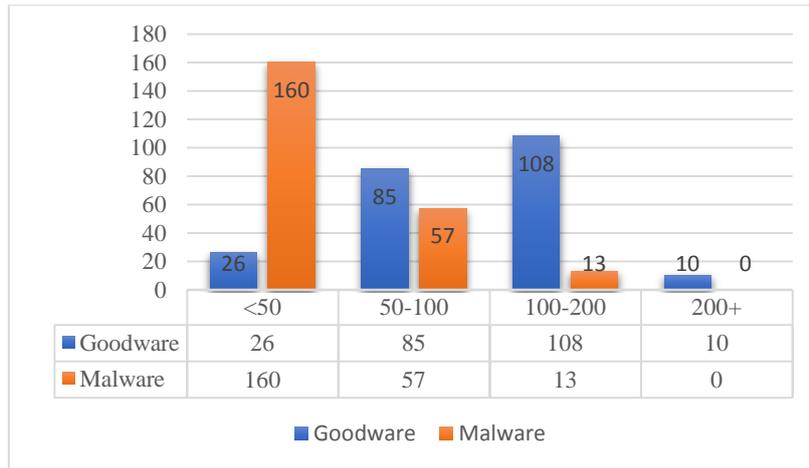

*Fig. 4 OpCode distribution*

The average density for the top 20 opcodes in each class is illustrated in Fig. 5.
It shows that the MOV opcode has the highest density at over 30% for both malware and goodware, compared to the second highest opcode, PUSH with around 13%. The densities for both classes are quite similar, especially for MOV and PUSH. Table 3 shows the top 20 opcodes for each class.

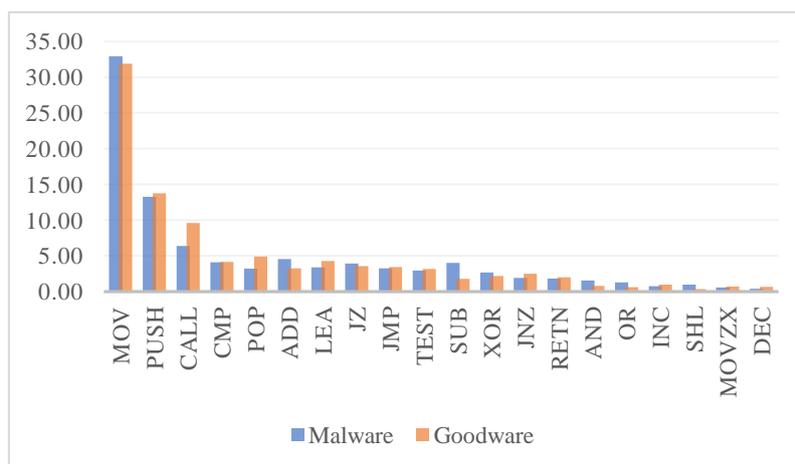

*Fig. 5 OpCode density*

*Table 3*
*Top 20 Opcode densities (%)*

|      | Goodware |         | Malware |         |      | Goodware |         | Malware |         |
|------|----------|---------|---------|---------|------|----------|---------|---------|---------|
| Rank | Opcode   | Density | Opcode  | Density | Rank | Opcode   | Density | Opcode  | Density |
| 1    | MOV      | 31.9%   | MOV     | 32.9%   | 11   | JNZ      | 2.5%    | TEST    | 2.9%    |
| 2    | PUSH     | 13.8%   | PUSH    | 13.2%   | 12   | XOR      | 2.2%    | XOR     | 2.7%    |
| 3    | CALL     | 9.6%    | CALL    | 6.4%    | 13   | RETN     | 2.0%    | JNZ     | 1.9%    |
| 4    | POP      | 4.9%    | ADD     | 4.5%    | 14   | SUB      | 1.8%    | RETN    | 1.8%    |
| 5    | LEA      | 4.3%    | CMP     | 4.1%    | 15   | INC      | 1.0%    | AND     | 1.6%    |
| 6    | CMP      | 4.2%    | SUB     | 4.0%    | 16   | AND      | 0.8%    | OR      | 1.3%    |
| 7    | JZ       | 3.5%    | JZ      | 3.9%    | 17   | MOVZX    | 0.7%    | SHL     | 1.0%    |
| 8    | JMP      | 3.4%    | LEA     | 3.4%    | 18   | DEC      | 0.7%    | INC     | 0.8%    |
| 9    | ADD      | 3.2%    | JMP     | 3.3%    | 19   | OR       | 0.6%    | SAR     | 0.7%    |
| 10   | TEST     | 3.2%    | POP     | 3.2%    | 20   | JLE      | 0.3%    | MOVZX   | 0.6%    |

### 3.4 Machine learning classification

The final phase involves the training and testing of the SVM classifier in the Waikato Environment for Knowledge Analysis 3.8.1 (WEKA) machine learning toolset [49]. WEKA was selected as the best tool for this experiment as it incorporates several standard machine learning techniques for simple workflow, can output results into a text and statistical format and offers several attribute selection evaluators.

#### 3.4.1 SVM and kernel functions

The Support Vector Machine (SVM) is a supervised machine learning technique, especially effective where binary classification is required; it is therefore an effective classifier for malware detection which often has two classes for prediction – malicious (malware) and benign (goodware). An SVM classifies data by constructing a hyperplane to separate data points in one class from those in another class; the best hyperplane therefore should have the largest margin of separation between both classes [50]. Where linear separation is not possible SVM can use kernel functions to transform the data into a higher dimensional feature space e.g. using by a radial hypersphere to achieve separation [51]. The goal of an SVM is therefore to produce a classification model based on a training dataset which is then used to predict the target values of the test dataset. [48].

Sequential Minimal Optimization, (or SMO), is an algorithm used to train Support Vector Machines, devised by John Platt in 1998. The SMO can handle large datasets and scales well due to the memory footprint growing linearly with the training set size.

In WEKA, the SMO classifier was run with all the default options including the selection of a 'Logistic' calibration method.

The SMO algorithm in WEKA uses linear, polynomial and Gaussian kernel functions and each kernel choice (Poly, NormalisedPoly, PUK, RBF) was initially tested using default values before further tuning and optimisation was applied.

*3.4.2   Feature / attribute selection process*

The attribute selection / feature reduction process is performed to enable the classifier to use the lowest number of features while maintaining the highest level of precision. Feature reduction applied to very large datasets can significantly decrease the training time for algorithms and the computational overhead associated with the high number of attributes and instances. Accuracy can also be increased by filtering out the noisy attributes that could have a negative effect on classification.

Attribute selection methods can be grouped into 2 categories: wrapper methods and filter methods. Wrapper methods can achieve better performance than filter methods due to their ability to be optimised, however they cannot always be used in datasets with a large number of attributes due to the evaluation process requiring a computational overhead and excessively long training times [52].

Within the WEKA toolset there are 9 attribute selection evaluators. The WrapperSubsetEval options is not used in this experiment as it was briefly tested and proved unresponsive for the reasons already discussed. The following attribute selection methods were evaluated: CfsSubsetEval**,** CorrelationAttributeEval, GainRatioAttributeEval, InfoGainAttribute
, OneRAttributeEval, Principal Component Analysis, ReliefFAttributeEval
, SymmetricalUncertAttributeEval.

WEKA has several search methods used by the attribute selection evaluators to perform either feature reduction, or ranking; in this experiment 3 search methods are employed: 1) 'BestFirst' uses greedy hill climbing (performing evaluation at each iteration) to search the attribute subspace; it can be configured to start with no attributes and search forwards, start with all attributes and search backwards, or search both directions from any point. It has a "backtracking" option which is used to terminate the search when the number of consecutive non-improving nodes is reached; 2) 'GreedyStepwise' performs a greedy search through the attribute subset space and can also be configured to start with no attributes and search forwards, all attributes and search backwards, or search both directions from any point. When the addition or removal of any remaining attributes results in a decrease in evaluation, the search will stop. It can be configured to generate a ranking by recording the order that the attributes are selected. Used in conjunction with the "numToSelect" option (specifies the number of attributes to retain and "threshold" option (threshold at which attributes are discarded) further feature reduction can be achieved; 3) 'Ranker' generates a ranking for each attribute based on their individual evaluations. Used in conjunction with the "numToSelect" option (specifies the number of attributes to retain and "threshold" option (threshold at which attributes are discarded) further feature reduction can be achieved.

3.5   Implementation

The experiment can be separated into 4 district phases:

1) Pre-processing the dataset

2) Creating the training and test datasets

3) Training and testing the SVM classifier
   Training and testing the attribute selection evaluators

4) Tuning the attribute selection evaluators to achieve further feature reduction

These phases are illustrated in workflow diagram Fig. 6.

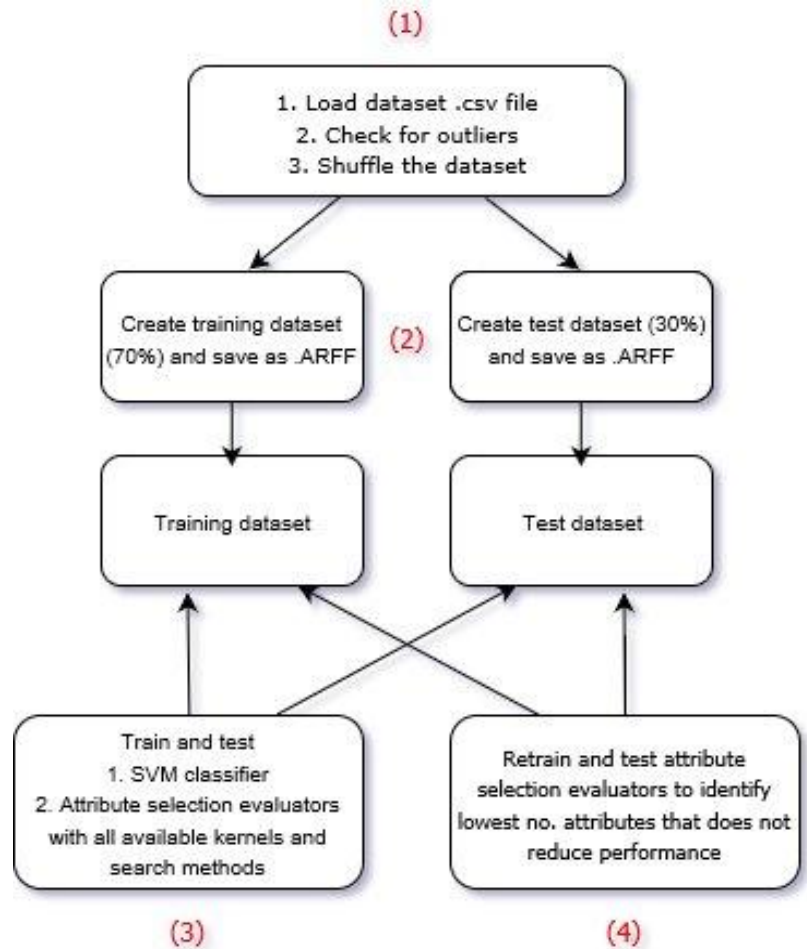

*Fig. 6 The 4 experiment phases*

.

### 3.5.1   Pre-processing the dataset (1)

The full dataset CSV file was loaded into the WEKA pre-processing tool to be prepared for the classifier tasks.

Outliers and extreme values are identified and removed in a two-step process.
In step 1 the '*weka.filters.unsupervised.attribute.InterquartileRange*' filter was applied to the loaded dataset to identify outliers and extreme values based on the results of interquartile aggregation.

In step 2 the corresponding '*weka.filters.unsupervised.instance.RemoveWithValues*' filter should be applied to remove the 'outlier' and 'ExtremeValue' attributes labelled as 'yes', however in this instance this step was not applied due to the high number of outliers detected. It was decided to continue with the experiment to see if the classification produced satisfactory results as Interquartile aggregation may not be a suitable method for outlier detection for the type of data in this dataset.

Finally the dataset was shuffled using the filter *weka.filters.unsupervised.instance.Randomize -S 42* to prevent an unequal sample distribution across either training, or test dataset.

### 3.5.2 Creating the training and test datasets (2)

Following shuffling the dataset was split into a training dataset (70%) and test (evaluation) dataset (30%). The following WEKA filters were applied and after each application the new dataset was saved in the. ARFF (Attribute-Relation File Format) format - a format consisting of a separate header and data section developed for use specifically within the WEKA software:

- *weka.filters.unsupervised.instance.RemovePercentage -P 30.0*
  (to remove 30% from each class in the dataset)

- *weka.filters.unsupervised.instance.RemovePercentage -P 30.0 -V*
  (selection inverted to swap the 30% back into the dataset)

The resulting training and test datasets contained 343 and 103 instances respectively.

### 3.5.3 Training and testing the SVM classifier (3.1)

The SVM classification experiment was conducted in multiple phases with each classification run using all available kernel and search methods.

The 10 main phases of the classification experiments are summarised in Table 4.

*Table 4*
*The 10 main phases of the classifications*

| Experiment | Options | Kernel Selections | Available Search methods |
|---|---|---|---|
| 1. SMO Classifier<br>2* Classes<br>(malware, goodware) | 1. Training and evaluation | - Poly<br>- NormalisedPoly<br>- PUK<br>- RBF | - |
| 2. SMO Classifier<br>6* Classes<br>(malware 1-5, goodware) | 1. Training and evaluation | - Poly<br>- NormalisedPoly<br>- PUK<br>- RBF | - |
| 3. CFSSubsetEval | 1. Attribute Selection<br>   Training and evaluation<br>2. Feature Reduction<br>   Training and evaluation | - Poly<br>- NormalisedPoly<br>- PUK<br>- RBF | BestFirst<br>GreedyStepwise |
| 4. CorrelationAttributeEval | 1. Attribute Selection<br>   Training and evaluation<br>2. Feature Reduction<br>   Training and evaluation | - Poly<br>- NormalisedPoly<br>- PUK<br>- RBF | Ranker |
| 5. GainRatioAttributeEval | 1. Attribute Selection<br>   Training and evaluation<br>2. Feature Reduction<br>   Training and evaluation | - Poly<br>- NormalisedPoly<br>- PUK<br>- RBF | Ranker |
| 6. InfoGainAttributeEval | 1. Attribute Selection<br>   Training and evaluation<br>2. Feature Reduction<br>   Training and evaluation | - Poly<br>- NormalisedPoly<br>- PUK<br>- RBF | Ranker |
| 7. OneRAttributeEval | 1. Attribute Selection<br>   Training and evaluation<br>2. Feature Reduction<br>   Training and evaluation | - Poly<br>- NormalisedPoly<br>- PUK<br>- RBF | Ranker |
| 8. PrincipalComponents<br>   - correlation matrix<br>   - covariance matrix | 1. Attribute Selection<br>   Training and evaluation<br>2. Feature Reduction<br>   Training and evaluation | - Poly<br>- NormalisedPoly<br>- PUK<br>- RBF | Ranker |
| 9. ReliefAttributeEval | 1. Attribute Selection<br>   Training and evaluation<br>2. Feature Reduction<br>   Training and evaluation | - Poly<br>- NormalisedPoly<br>- PUK<br>- RBF | Ranker |
| 10. SymmetricalUncertAttributeEval | 1. Attribute Selection<br>   Training and evaluation<br>2. Feature Reduction<br>   Training and evaluation | - Poly<br>- NormalisedPoly<br>- PUK<br>- RBF | Ranker |

In each of the classification experiments training was performed using 10-fold cross validation. Cross-validation is an important technique used to help prevent overfitting by allocating a portion of the training dataset for validation, validating using multiple rounds (the number of rounds is determined by the number of specified folds, in this case 10) and then averaging the prediction results. Using 10-folds has become a standard process for cross-validation and following extensive testing it has been shown that around this number achieves the best estimate of error [53]. All SMO classifier options were set to their default values and the experiment repeated for each of the 4 available kernel functions (Poly, NormalisedPoly, PUK, RBF)

The test dataset (containing 30%, 103 instances) was supplied to the classifier and each model re-evaluated against the same test set. The evaluation metrics are compared in the following result tables. The highest performing evaluated model is highlighted in green, the lowest in red.

### 3.5.4 Training and testing the attribute selection evaluators (3.2)

In Phases 3 - 10 the Attribute Selection evaluators were run using default options and a single parameter changed to discard attributes below a threshold value of 0; this was selected to achieve an initial feature reduction prior to further tuning.
The dimensionality of the data was therefore reduced by the attribute selection evaluation process before being passed onto the selected classifier, in this case, SMO.
Classifications were performed using all available search methods and kernels (using default, untuned settings) and the results presented.

### 3.5.5 Evaluation metrics

The performance of each classification model was evaluated using the 5 default WEKA metrics as summarised in Table 6 and compared in the later Experiments and Results section.

*Table 5*
*Evaluation metrics* [54]

| Metric | Calculation | Value |
|---|---|---|
| True positive rate (TPR) | *TP/(TP+FN)* | Rate of true positives (instances correctly classified as a given class) |
| False positive rate (FPR) | *FP/(FP+TN)* | Rate of false positives (instances falsely classified as a given class) |
| Precision | *TP/(TP+FP)* | Proportion of instances that are truly of a class divided by the total instances classified as that class |
| Recall | *TPR* | Proportion of instances classified as a given class divided by the actual total in that class (equivalent to TP rate) |
| F-measure | *2 × (Recall × Precision)/ (Recall + Precision)* | A combined measure for precision and recall (Estimate of entire system performance) |

TP= True positive, FN=false negative, TN= true negative, FP=false positive

### 3.5.6 Machine specifications

Best practice within malware analysis and malicious executable handling is to use an isolated virtual environment that is detached from its host. This provides the ability to restore the environment to a clean, previous state, to recover from, or clean-up after infection [44], [55]. The feature extraction phase of the experiment was undertaken within an isolated virtual environment on a lab PC with the network cable disconnected to further isolate it from the network. The virtual environment used for data collection and feature extraction was VMWare Workstation due to the ease of configuration, deployment and familiarity as the environment was already configured on the host laptop with a number of suitable VMs already in use for such tasks. Table 7 outlines the environment machine specifications for each stage of the experiment.

*Table 6*
*Machine specifications for each stage of the experiment*

| Phase | Host | Virtual environment | Application |
|---|---|---|---|
| 1. Data Collection | Viglen Genie Desktop PC:<br>• Windows 7 Enterprise SP1<br>• 8GB RAM<br>• Intel Core 2 Quad CPU Q8400@ 2.66GHz | - VMWorkstation 12 Pro 12.1.1 build 3770994<br><br>- Debian 4.9.6-3 Kali2 (2017-01-30) i686 GNU/Linux<br>3GB RAM<br>4 processors<br><br>- NAT mode<br>(used to share the hosts ip address) | Mozilla Firefox ESR 45.7.0 |
| 2. Feature Extraction | | - VMWorkstation 12 Pro 12.1.1 build 3770994<br><br>- Windows 7 Professional SP1<br>1GB RAM<br>1 Processor<br><br>- Host only mode<br>( a private network shared with the host only) | IDA 6.95.160926 (32-bit) |
| 3. Dataset creation | Lenovo IdeaPad 310<br>• Windows 10 Home<br>• 8GB RAM<br>• Intel Core i5 6200U @2.40GHz | - | Microsoft Excel 2016 MSO (16.1.7726.1049) 64 bit |
| 4. Machine Learning Classifier | | - | WEKA.version 3.8.1 Java(TM) SE Runtime Environment 1.8.0_112-b15 java.version 1.8.0_112 |

## 4. Experiments and Results

### 4.1 SMO (2 classes)

In Phase 1 the SMO classifier was trained and tested using the classifier default options to obtain a benchmark set of evaluation metrics for all extracted attributes in the detection model. (Table 8). The 2 classes used were 'malware' and 'goodware'.

*Table 7*
*SMO (2classes) evaluation metrics*

| Model | Kernel | Action | TPR | FPR | Precision | Recall | F-Measure | Attributes |
|---|---|---|---|---|---|---|---|---|
| 1 | Poly [1] | Training (10-fold cross-validation) | 90.7% | 9.5% | 90.7% | 90.7% | 90.7% | 444 |
| 1 | Poly [1] | Re-evaluation against test set | 94.2% | 5.9% | 94.4% | 94.2% | 94.2% | 444 |
| 2 | NormalisedPoly [2] | Training (10-fold cross-validation) | 89.8% | 10.7% | 90.0% | 89.8% | 89.8% | 444 |
| 2 | NormalisedPoly [2] | Re-evaluation against test set | 94.2% | 5.9% | 94.4% | 94.2% | 94.2% | 444 |
| 3 | PUK | Training (10-fold cross-validation) | 86.9% | 14.0% | 87.5% | 86.9% | 86.8% | 444 |
| 3 | PUK | Re-evaluation against test set | 100.0% | 0.0% | 100.0% | 100.0% | 100.0% | 444 |
| 4 | RBF | Training (10-fold cross-validation) | 80.5% | 20.2% | 80.6% | 80.5% | 80.4% | 444 |
| 4 | RBF | Re-evaluation against test set | 82.5% | 17.4% | 82.6% | 82.5% | 82.5% | 444 |

[1] Exponent default value set to 1.0 to achieve linear kernel function
[2] Exponent default value set to 2.0 to achieve quadratic kernel function

With all default options set the PUK kernel (Pearson VII function-based universal kernel) achieved the best results with 100% precision and a 0% False Positive rate (FPR).

The linear and quadratic kernel functions achieved slightly less than the target 95% precision and the RBF Gaussian kernel function only achieving 82.6% precision.

Determining the best kernel function for the dataset can often only be achieved by conducting multiple experiments across all available kernels and parameters, and comparing the chosen evaluation metrics therefore the Poly, NormalisedPoly and RBF kernels were further tuned to achieve higher precision comparable to the PUK kernel results:

- Poly/NormalisedPoly:
'Complexity' tested using values 0.1, 1.0, 10.0, 100.0; 'exponent using' 1.0, 2.0, 3.0

- RBF kernel:
'Complexity' tested using values 0.1, 1.0, 10.0, 100.0; 'gamma' using 0.01, 0.1, 1.0, 10.0

Table 9 below summarises the parameter selections that provided the highest precision.

*Table 8*
*New parameter selections for tuning the SMO kernels*

| Kernel | Changed parameters | Action | TPR | FPR | Precision | Recall | F-Measure | Attributes |
|---|---|---|---|---|---|---|---|---|
| Poly | E = 2 C = 1 | Training (10-fold cross-validation) | 90.4% | 9.6% | 90.4% | 90.4% | 90.4% | 444 |
| Poly | E = 2 C = 1 | Re-evaluation against test set | 100.0% | 0.0% | 100.0% | 100.0% | 100.0% | 444 |
| Poly | E = 1 C = 10 | Training (10-fold cross-validation) | 90.4% | 9.6% | 90.4% | 90.4% | 90.4% | 444 |
| Poly | E = 1 C = 10 | Re-evaluation against test set | 100.0% | 0.0% | 100.0% | 100.0% | 100.0% | 444 |
| NormalisedPoly | E = 2 C = 100 | Training (10-fold cross-validation) | 89.8% | 10.5% | 89.8% | 89.8% | 89.8% | 444 |
| NormalisedPoly | E = 2 C = 100 | Re-evaluation against test set | 100.0% | 0.0% | 100.0% | 100.0% | 100.0% | 444 |
| NormalisedPoly | E = 3 C = 10 | Training (10-fold cross-validation) | 91.5% | 8.8% | 91.6% | 91.5% | 91.5% | 444 |
| NormalisedPoly | E = 3 C = 10 | Re-evaluation against test set | 100.0% | 0.0% | 100.0% | 100.0% | 100.0% | 444 |
| RBF | γ = 1 C = 10 | Training (10-fold cross-validation) | 84.5% | 16.4% | 85.1% | 84.5% | 84.4% | 444 |
| RBF | γ = 1 C = 10 | Re-evaluation against test set | 100.0% | 0.0% | 100.0% | 100.0% | 100.0% | 444 |

C -- The complexity parameter.
E -- The exponent value
γ -- The Gamma value.

Each kernel, when properly tuned, can achieve 100% precision when re-evaluated against the test dataset.

### 4.2     SMO (6 classes)

Phase 2 required a second version of the dataset to be created and labelled with 6 different classes to represent the benign instances and 5 different ransomware classes. (A single dataset with 2 class attributes could have been used but creating separate datasets reduced the complexity) The steps in 3.5.1 and 3.5.2 were repeated to pre-process the data and create the training and test datasets for use in the 6-class classification model.

*Table 9*
*SMO (6classes) evaluation metrics*

| Model | Kernel | Action | TPR | FPR | Precision | Recall | F-Measure | Attributes |
|---|---|---|---|---|---|---|---|---|
| 45 | Poly [1] | Training (10-fold cross-validation) | 70.8% | 10.2% | 70.0% | 70.8% | 69.8% | 444 |
| 45 | Poly [1] | Re-evaluation against test set | 87.4% | 4.1% | 86.7% | 87.4% | 86.7% | 444 |
| 46 | NormalisedPoly [2] | Training (10-fold cross-validation) | 69.4% | 14.6% | 66.9% | 69.4% | 66.0% | 444 |
| 46 | NormalisedPoly [2] | Re-evaluation against test set | 87.4% | 6.7% | 86.3% | 87.4% | 86.6% | 444 |
| 47 | PUK | Training (10-fold cross-validation) | 62.7% | 26.2% | 60.9% | 62.7% | 56.0% | 444 |
| 47 | PUK | Re-evaluation against test set | 97.1% | 0.3% | 96.5% | 97.1% | 96.7% | 444 |
| 48 | RBF | Training (10-fold cross-validation) | 49.3% | 43.9% | 35.1% | 49.3% | 34.7% | 444 |
| 48 | RBF | Re-evaluation against test set | 51.5% | 47.6% | 46.4% | 51.5% | 36.8% | 444 |

[1] Exponent default value set to 1.0 to achieve linear kernel function
[2] Exponent default value set to 2.0 to achieve quadratic kernel function

The evaluation metrics for this phase of the experiment can be seen in Table 10.
With all default options set the PUK kernel (Pearson VII function-based universal kernel) once again achieved the best results with 96.5% precision and 0.3% False Positive rate (FPR). Table 11 shows the detailed accuracy by each class in the classification. Three classes

(Good, TeslaCrypt, Cryptowall) achieved 100% precision and although Cerber achieved a 100% TPR, 2 Locky samples were incorrectly detected as Cerber resulting in a decrease in precision due to receiving false positives. The confusion matrix for this association can be seen in Table 12. Although all classes were represented in the training dataset, the test dataset did not contain any instances for the Torrentlocker class which highlights an issue in that this class was not represented correctly in the dataset.

*Table 10*
*Detailed accuracy by class (SMO 6 classes)*

| TPR | FPR | Precision | Recall | F-measure | Class |
|---|---|---|---|---|---|
| 100.0% | 0.0% | 100.0% | 100.0% | 100.0% | Good |
| 0.0% | 0.0% | 0.0% | 0.0% | 0.0% | Torrentlocker |
| 100.0% | 0.0% | 100.0% | 100.0% | 100.0% | Teslacrypt |
| 86.7% | 1.1% | 92.9% | 86.7% | 89.7% | Locky |
| 100.0% | 0.0% | 100.0% | 100.0% | 100.0% | Cryptowall |
| 100.0% | 2.1% | 75.0% | 100.0% | 85.7% | Cerber |
| 97.1% | 0.3% | 96.5% | 97.1% | 96.6% | Weighted avg. |

*Table 11*
*Confusion matrix with false positives highlighted in red (SMO 6 classes)*

| a | b | c | d | e | f | <-- classified as |
|---|---|---|---|---|---|---|
| 51 | 0 | 0 | 0 | 0 | 0 | \| a = good |
| 0 | 0 | 0 | 1 | 0 | 0 | \| b = Torrentlocker |
| 0 | 0 | 22 | 0 | 0 | 0 | \| c = TeslaCrypt |
| 0 | 0 | 0 | 13 | 0 | 2 | \| d = Locky |
| 0 | 0 | 0 | 0 | 8 | 0 | \| e = CryptoWall |
| 0 | 0 | 0 | 0 | 0 | 6 | \| f = Cerber |

## 4.3 Training and testing the attribute selection evaluators (3.2)

In Phases 3 - 10 the Attribute Selection evaluators were run using default options and a single parameter changed to discard attributes below a threshold value of 0; this was selected to achieve an initial feature reduction prior to further tuning.
The dimensionality of the data was therefore reduced by the attribute selection evaluation process before being passed onto the selected classifier, in this case, SMO.
Classifications were performed using all available search methods and kernels (using default, untuned settings) and the results presented.

### 4.3.1 CFSSubsetEval

CfsSubsetEval is a Subset evaluator that is used conjunction with an appropriate search option to determine the smallest subset size that has the same consistency as the full
attribute set [53]. It assesses the predictive ability of each individual attribute and the degree of redundancy among them; it prefers sets of attributes that are highly-correlated with the class, but have low intercorrelation with each other. The CfsSubsetEval evaluator determined that a subset of 21 attributes had a consistency equal to the full set of 444 attributes.

The highest performing model using the PUK kernel achieved 94.2% precision using 21 attributes. (Table 13) The ranked 21 attributes are: SETS, SETNBE, SETNLE, JB, FSUB, XCHG, POP, OR, FUCOMPP, JLE, CMOVS, ROR, FIDIV, SETBE, JA, LEA, FNINIT, CALL, AND, SETLE, FDIVP

*Table 12*
*CfsSubsetEval evaluation metrics*

| Model | Search method | Kernel | Action | TPR | FPR | Precision | Recall | F-Measure | Attributes |
|---|---|---|---|---|---|---|---|---|---|
| 5 | BestFirst | Poly | Training (10-fold cross-validation) | 87.2% | 13.7% | 87.7% | 87.2% | 87.1% | 21 |
| 5 | BestFirst | Poly | Re-evaluation against test set | 87.4% | 12.7% | 87.7% | 87.4% | 87.3% | 21 |
| 6 | GreedyStepwise | Poly | Training (10-fold cross-validation) | 87.2% | 13.6% | 87.6% | 87.2% | 87.1% | 21 |
| 6 | GreedyStepwise | Poly | Re-evaluation against test set | 87.4% | 12.7% | 87.7% | 87.4% | 87.3% | 21 |
| 7 | BestFirst | NormalisedPoly | Training (10-fold cross-validation) | 90.1% | 10.3% | 90.2% | 90.1% | 90.1% | 21 |
| 7 | BestFirst | NormalisedPoly | Re-evaluation against test set | 87.4% | 12.6% | 87.4% | 87.4% | 87.4% | 21 |
| 8 | GreedyStepwise | NormalisedPoly | Training (10-fold cross-validation) | 90.1% | 10.3% | 90.1% | 90.1% | 90.1% | 21 |
| 8 | GreedyStepwise | NormalisedPoly | Re-evaluation against test set | 87.4% | 12.6% | 87.4% | 87.4% | 87.4% | 21 |
| 9 | BestFirst | PUK | Training (10-fold cross-validation) | 92.1% | 8.1% | 92.2% | 92.1% | 92.1% | 21 |
| 9 | BestFirst | PUK | Re-evaluation against test set | 93.2% | 6.8% | 93.2% | 93.2% | 93.2% | 21 |
| 10 | GreedyStepwise | PUK | Training (10-fold cross-validation) | 92.1% | 8.1% | 92.2% | 92.1% | 92.1% | 21 |
| 10 | GreedyStepwise | PUK | Re-evaluation against test set | 94.2% | 5.9% | 94.2% | 94.2% | 94.2% | 21 |
| 11 | BestFirst | RBF | Training (10-fold cross-validation) | 54.2% | 51.7% | 75.4% | 54.2% | 39.3% | 21 |
| 11 | BestFirst | RBF | Re-evaluation against test set | 51.5% | 49.5% | 75.3% | 51.5% | 36.0% | 21 |
| 12 | GreedyStepwise | RBF | Training (10-fold cross-validation) | 54.2% | 51.7% | 75.4% | 54.2% | 39.3% | 21 |
| 12 | GreedyStepwise | RBF | Re-evaluation against test set | 51.5% | 49.5% | 75.3% | 51.5% | 36.0% | 21 |

### 4.3.2 CorrelationAttributeEval

CorrelationAttributeEval is another Correlation-based evaluator but it uses ranking to evaluate each attribute and does not apply filtering. This is used in conjunction with the "discard" and "threshold" options to achieve feature reduction.

The CorrelationAttributeEval evaluator uses the ranker search method so only achieved a low initial feature reduction due to the zero-discard threshold setting. It achieved 100% precision with the PUK kernel and this model was selected for further tuning. (Table 14)

*Table 13*
*CorrelationAttributeEval evaluation metrics*

| Model | Search method | Kernel | Action | TPR | FPR | Precision | Recall | F-Measure | Attributes |
|---|---|---|---|---|---|---|---|---|---|
| 13 | Ranker | Poly | Training (10-fold cross-validation) | 90.7% | 9.5% | 90.7% | 90.7% | 90.7% | 432 |
| 13 | Ranker | Poly | Re-evaluation against test set | 94.2% | 5.9% | 94.4% | 94.2% | 94.2% | 432 |
| 14 | Ranker | NormalisedPoly | Training (10-fold cross-validation) | 89.8% | 10.7% | 90.0% | 89.8% | 89.8% | 432 |
| 14 | Ranker | NormalisedPoly | Re-evaluation against test set | 94.2% | 5.9% | 94.4% | 94.2% | 94.2% | 432 |
| 15 | Ranker | PUK | Training (10-fold cross-validation) | 86.9% | 14.0% | 87.5% | 86.9% | 86.8% | 432 |
| 15 | Ranker | PUK | Re-evaluation against test set | 100.0% | 0.0% | 100.0% | 100.0% | 100.0% | 432 |

| Model | Search method | Kernel | Action | TPR | FPR | Precision | Recall | F-Measure | Attributes |
|---|---|---|---|---|---|---|---|---|---|
| 16 | Ranker | RBF | Training (10-fold cross-validation) | 80.2% | 20.5% | 80.3% | 80.2% | 80.1% | 432 |
| 16 | Ranker | RBF | Re-evaluation against test set | 82.5% | 17.4% | 82.6% | 82.5% | 82.5% | 432 |

### 4.3.3 GainRatioAttributeEval

The GainRatioAttributeEval evaluator measures the gain ratio with respect to the class to evaluate predictor importance of each attribute [54]. Using the zero-discard threshold a feature reduction from 444 to 252 attributes and 99% precision using the PUK kernel was achieved. (Table 15)

Table 14
*GainRatioAttributeEval evaluation metrics*

| Model | Search method | Kernel | Action | TPR | FPR | Precision | Recall | F-Measure | Attributes |
|---|---|---|---|---|---|---|---|---|---|
| 17 | Ranker | Poly | Training (10-fold cross-validation) | 91.3% | 8.9% | 91.3% | 91.3% | 91.3% | 252 |
| 17 | Ranker | Poly | Re-evaluation against test set | 94.2% | 5.9% | 94.4% | 94.2% | 94.2% | 252 |
| 18 | Ranker | NormalisedPoly | Training (10-fold cross-validation) | 91.0% | 9.3% | 91.0% | 91.0% | 91.0% | 252 |
| 18 | Ranker | NormalisedPoly | Re-evaluation against test set | 95.1% | 4.9% | 95.3% | 95.1% | 95.1% | 252 |
| 19 | Ranker | PUK | Training (10-fold cross-validation) | 90.4% | 9.7% | 90.4% | 90.4% | 90.4% | 252 |
| 19 | Ranker | PUK | Re-evaluation against test set | 99.0% | 1.0% | 99.0% | 99.0% | 99.0% | 252 |
| 20 | Ranker | RBF | Training (10-fold cross-validation) | 81.0% | 19.5% | 81.1% | 81.0% | 81.0% | 252 |
| 20 | Ranker | RBF | Re-evaluation against test set | 82.5% | 17.4% | 82.6% | 82.5% | 82.5% | 252 |

### 4.3.4 InfoGainAttributeEval

The InfoGainAttribute evaluator measures the information gain with respect to the class to evaluate predictor importance of each attribute [54]. Using the zero-discard threshold the same results as the GainRatioAttributeEval evaluator were achieved i.e. feature reduction from 444 to 252 attributes and 99% precision using the PUK kernel. (Table 16)

Table 15
*InfoGainAttribute evaluation metrics*

| Model | Search method | Kernel | Action | TPR | FPR | Precision | Recall | F-Measure | Attributes |
|---|---|---|---|---|---|---|---|---|---|
| 21 | Ranker | Poly | Training (10-fold cross-validation) | 91.3% | 8.9% | 91.3% | 91.3% | 91.3% | 252 |
| 21 | Ranker | Poly | Re-evaluation against test set | 94.2% | 5.9% | 94.4% | 94.2% | 94.2% | 252 |
| 22 | Ranker | NormalisedPoly | Training (10-fold cross-validation) | 91.0% | 9.3% | 91.0% | 91.0% | 91.0% | 252 |
| 22 | Ranker | NormalisedPoly | Re-evaluation against test set | 95.1% | 4.9% | 95.3% | 95.1% | 95.1% | 252 |
| 23 | Ranker | PUK | Training (10-fold cross-validation) | 90.4% | 9.7% | 90.4% | 90.4% | 90.4% | 252 |
| 23 | Ranker | PUK | Re-evaluation against test set | 99.0% | 1.0% | 99.0% | 99.0% | 99.0% | 252 |
| 24 | Ranker | RBF | Training (10-fold cross-validation) | 81.0% | 19.5% | 81.1% | 81.0% | 81.0% | 252 |
| 24 | Ranker | RBF | Re-evaluation against test set | 82.5% | 17.4% | 82.6% | 82.5% | 82.5% | 252 |

### 4.3.5 OneRAttributeEval

OneRAttributeEval evaluates the worth of an attribute by using the OneR classifier, a simple cheap classifier that can obtain high accuracy using of a set of rules that all test one particular attribute to determine the class of an instance and aims to find the attribute with the fewest prediction errors [56].

Using the zero-discard threshold the evaluator initially performed only a single attribute feature reduction by removing the 'class' attribute, however it achieved 100% precision using the PUK kernel. (Table 17)

*Table 16*
*OneRAttributeEval evaluation metrics*

| Model | Search method | Kernel | Action | TPR | FPR | Precision | Recall | F-Measure | Attributes |
|---|---|---|---|---|---|---|---|---|---|
| 25 | Ranker | Poly | Training (10-fold cross-validation) | 90.7% | 9.5% | 90.7% | 90.7% | 90.7% | 443 |
| 25 | Ranker | Poly | Re-evaluation against test set | 94.2% | 5.9% | 94.4% | 94.2% | 94.2% | 443 |
| 26 | Ranker | NormalisedPoly | Training (10-fold cross-validation) | 89.8% | 10.7% | 90.0% | 89.8% | 89.8% | 443 |
| 26 | Ranker | NormalisedPoly | Re-evaluation against test set | 94.2% | 5.9% | 94.4% | 94.2% | 94.2% | 443 |
| 27 | Ranker | PUK | Training (10-fold cross-validation) | 86.9% | 14.0% | 87.5% | 86.9% | 86.8% | 443 |
| 27 | Ranker | PUK | Re-evaluation against test set | 100.0% | 0.0% | 100.0% | 100.0% | 100.0% | 443 |
| 28 | Ranker | RBF | Training (10-fold cross-validation) | 80.2% | 20.5% | 80.3% | 80.2% | 80.1% | 443 |
| 28 | Ranker | RBF | Re-evaluation against test set | 82.5% | 17.4% | 82.6% | 82.5% | 82.5% | 443 |

### 4.3.6 PrincipalComponents

Principal Component Analysis is a widely-used dimension reduction technique. It is based on the principle of converting a large number of variables into a smaller number of uncorrelated variables; it can reduce training and testing times for SVM classifiers with little decrease in accuracy [57]. The new attributes are then ranked in order of their eigenvalues with a subset of attributes selected by choosing sufficient eigenvectors to account for a specified proportion of the variance, usually 95%. The attributes can then be transformed back to their original space but with a loss of ranking for predictor importance.

The PCA evaluation was performed using both the covariance and correlation (to perform standardisation) matrices which gave very different results.

When run with the default values the covariance matrix achieved significantly higher dimension reduction resulting in just 7 attributes, but at the expense of performance. The highest precision using the correlation matrix and PUK kernel achieved just under the target 95% precision with a value of 94.4%, with a reduction in attributes to 87. (Table 18)

*Table 17*
*PCA evaluation metrics*

| Model | Search method | Kernel | Action | TPR | FPR | Precision | Recall | F-Measure | Attributes |
|---|---|---|---|---|---|---|---|---|---|
| 29 [1] | Ranker | Poly | Training (10-fold cross-validation) | 85.4% | 15.5% | 85.9% | 85.4% | 85.3% | 87 |
| 29 [1] | Ranker | Poly | Re-evaluation against test set | 90.3% | 9.8% | 90.8% | 90.3% | 90.3% | 87 |
| 30 [1] | Ranker | NormalisedPoly | Training (10-fold cross-validation) | 54.2% | 51.6% | 63.7% | 54.2% | 40.3% | 87 |
| 30 [1] | Ranker | NormalisedPoly | Re-evaluation against test set | 50.5% | 50.5% | 25.5% | 50.5% | 33.9% | 87 |
| 31 [1] | Ranker | PUK | Training (10-fold cross-validation) | 87.2% | 12.5% | 87.4% | 87.2% | 87.2% | 87 |
| 31 [1] | Ranker | PUK | Re-evaluation against test set | 94.2% | 5.9% | 94.4% | 94.2% | 94.2% | 87 |
| 32 [1] | Ranker | RBF | Training (10-fold cross-validation) | 53.1% | 53.1% | 28.2% | 53.1% | 36.8% | 87 |
| 32 [1] | Ranker | RBF | Re-evaluation against test set | 50.5% | 50.5% | 25.5% | 50.5% | 33.9% | 87 |
| 33 [2] | Ranker | Poly | Training (10-fold cross-validation) | 72.9% | 29.6% | 76.2% | 72.9% | 71.5% | 7 |
| 33 [2] | Ranker | Poly | Re-evaluation against test set | 71.8% | 28.6% | 78.5% | 71.8% | 70.0% | 7 |
| 34 [2] | Ranker | NormalisedPoly | Training (10-fold cross-validation) | 54.2% | 51.6% | 63.7% | 54.2% | 40.3% | 7 |
| 34 [2] | Ranker | NormalisedPoly | Re-evaluation against test set | 50.5% | 50.5% | 25.5% | 50.5% | 33.9% | 7 |
| 35 [2] | Ranker | PUK | Training (10-fold cross-validation) | 85.1% | 14.3% | 85.7% | 85.1% | 85.1% | 7 |
| 35 [2] | Ranker | PUK | Re-evaluation against test set | 85.1% | 14.3% | 85.7% | 85.1% | 85.1% | 7 |
| 36 [2] | Ranker | RBF | Training (10-fold cross-validation) | 53.1% | 53.1% | 28.2% | 53.1% | 36.8% | 7 |
| 36 [2] | Ranker | RBF | Re-evaluation against test set | 50.5% | 50.5% | 25.5% | 50.5% | 33.9% | 7 |

[1] using the correlation matrix
[2] using the covariance matrix

### 4.3.7 RelieffAttributeEval

ReliefFAttributeEval is instance-based, sampling instances randomly and checking neighbouring instances of the same and different classes. It can operate on both discrete and continuous class data. [53]

The ReliefFAttributeEval, an instance-based evaluator achieved 99% precision using the PUK kernel, with a feature reduction to 281 features. (Table 19)

*Table 18*
*ReliefFAttributeEval evaluation metrics*

| Model | Search method | Kernel | Action | TPR | FPR | Precision | Recall | F-Measure | Attributes |
|---|---|---|---|---|---|---|---|---|---|
| 37 | Ranker | Poly | Training (10-fold cross-validation) | 91.0% | 9.2% | 91.0% | 91.0% | 91.0% | 281 |
| 37 | Ranker | Poly | Re-evaluation against test set | 94.2% | 5.9% | 94.4% | 94.2% | 94.2% | 281 |
| 38 | Ranker | NormalisedPoly | Training (10-fold cross-validation) | 90.7% | 9.8% | 90.8% | 90.7% | 90.6% | 281 |
| 38 | Ranker | NormalisedPoly | Re-evaluation against test set | 93.2% | 6.9% | 93.3% | 93.2% | 93.2% | 281 |
| 39 | Ranker | PUK | Training (10-fold cross-validation) | 90.4% | 10.2% | 90.6% | 90.4% | 90.3% | 281 |
| 39 | Ranker | PUK | Re-evaluation against test set | 99.0% | 1.0% | 99.0% | 99.0% | 99.0% | 281 |
| 40 | Ranker | RBF | Training (10-fold cross-validation) | 80.8% | 19.8% | 80.8% | 80.8% | 80.7% | 281 |
| 40 | Ranker | RBF | Re-evaluation against test set | 82.5% | 17.4% | 82.6% | 82.5% | 82.5% | 281 |

### 4.3.8 SymmetricalUncertAttributeEval

SymmetricalUncertAttributeEval evaluates the worth of an attribute by measuring the symmetrical uncertainty with respect to the class, assigning a value of 0 or 1 to represent irrelevance and relevance respectively [54]. It achieved very high precision, 99% with a feature reduction to 252 features. (Table 20)

*Table 19*
*SymmetricalUncertAttributeEval evaluation metrics*

| Desc | Search method | Kernel | Action | TPR | FPR | Precision | Recall | F-Measure | Attributes |
|---|---|---|---|---|---|---|---|---|---|
| 41 | Ranker | Poly | Training (10-fold cross-validation) | 91.3% | 8.9% | 91.3% | 91.3% | 91.3% | 252 |
| 41 | Ranker | Poly | Re-evaluation against test set | 94.2% | 5.9% | 94.4% | 94.2% | 94.2% | 252 |
| 42 | Ranker | NormalisedPoly | Training (10-fold cross-validation) | 91.0% | 9.3% | 91.0% | 91.0% | 91.0% | 252 |
| 42 | Ranker | NormalisedPoly | Re-evaluation against test set | 95.1% | 4.9% | 95.3% | 95.1% | 95.1% | 252 |
| 43 | Ranker | PUK | Training (10-fold cross-validation) | 90.4% | 9.7% | 90.4% | 90.4% | 90.4% | 252 |
| 43 | Ranker | PUK | Re-evaluation against test set | 99.0% | 1.0% | 99.0% | 99.0% | 99.0% | 252 |
| 44 | Ranker | RBF | Training (10-fold cross-validation) | 81.0% | 19.5% | 81.1% | 81.0% | 81.0% | 252 |
| 44 | Ranker | RBF | Re-evaluation against test set | 82.5% | 17.4% | 82.6% | 82.5% | 82.5% | 252 |

### 4.4 Tuning the attribute selection evaluators to achieve further feature reduction (4)

The best performing model from each phase model was tuned by applying a decremental change to the discard threshold within the Ranker or greedy stepwise searches to achieve increased feature reduction, but without a decrease in precision. Once a decrease had been reached, the highest number of attributes required to maintain performance was recorded.

In each of the classification experiments training was performed using 10-fold cross validation with each model re-evaluated against the same test set. Table 21 compares the final feature reduction achieved by each tuning process. The CorrelationAttributeEval still achieves 100% precision and 0% TPR with a significantly reduced no. of features (reduced to 180). CFSSubset is the lowest performing evaluator with a 94.2% precision rate but it does provide the highest feature reduction at only 10 features or a 97.7% reduction.

*Table 20*
*Comparison of feature reduction*

| Model | Desc | Search method | Kernel | TPR | FPR | Precision | Recall | F-Measure | Feature Reduction | Reduction |
|---|---|---|---|---|---|---|---|---|---|---|
| *3* | SMO (Benchmark) | - | *PUK* | *100.0%* | *0.0%* | *100.0%* | *100.0%* | *100.0%* | - | |
| *10* | CFSSubset | GreedyStepwise | *PUK* | 94.2% | 5.9% | 94.2% | 94.2% | 94.2% | *21 -> 10* | 97.7% |
| *15* | CorrelationAttributeEval | Ranker | *PUK* | 100.0% | 0.0% | 100.0% | 100.0% | 100.0% | *432 -> 180* | 59.5% |
| *19* | GainRatioAttributeEval | Ranker | *PUK* | 99.0% | 1.0% | 99.0% | 99.0% | 99.0% | *252 -> 183* | 58.8% |
| *23* | InfoGainAttributeEval | Ranker | *PUK* | 99.0% | 1.0% | 99.0% | 99.0% | 99.0% | *252 -> 128* | 71.2% |
| *27* | OneRAttributeEval | Ranker | *PUK* | 100.0% | 0.0% | 100.0% | 100.0% | 100.0% | *443 -> 332* | 25.2% |
| *31* | PrincipalComponents | Ranker | *PUK* | 94.2% | 5.9% | 94.4% | 94.2% | 94.2% | *87 -> 55* | 87.6% |
| *39* | ReliefFAttributeEval | Ranker | *PUK* | 99.0% | 1.0% | 99.0% | 99.0% | 99.0% | *281 -> 77* | 82.7%% |
| *43* | SymmetricalUncertAttributeEval | Ranker | *PUK* | 99.0% | 1.0% | 99.0% | 99.0% | 99.0% | *252 -> 133* | 70.0% |

## 4.5 Important Opcodes

Table 5-16 shows a list of the top 21 ranked Opcodes by each attribute selection evaluator. The top 21 values have been chosen to reflect the value (21) of the initial feature reduction achieved by the CFSSubsetEval method for full comparison. Due to the difficulty in calculating the specific ranked Opcodes by the PCA filter the attributes have been transformed back into their original space resulting in an equal rank being assigned to each attribute. The PCA top21 Opcodes have included on Table 22 for reference but are not ranked.

*Table 21*
*Top 21 ranked opcodes by feature reduction method*

| Rank | CFSSubset Evaluator | CorrelationAttribute Eval | GainRatioAttribute Eval | InfoGainAttribute Eval | OneRAttribute Eval | ReliefFAttribute Eval | Symmetrical UncertAttribute Eval | Principal Components [1] |
|---|---|---|---|---|---|---|---|---|
| 1 | SETS | JLE | SETLE | FDIVP | FDIVP | POP | FDIVP | SUBPS |
| 2 | SETNBE | CALL | FDIVP | XCHG | AND | INC | SETLE | MULPS |
| 3 | SETNLE | FILD | SETBE | AND | XCHG | JNZ | FILD | SETNBE |
| 4 | JB | POP | SETNBE | JA | SETLE | CMP | FSUBRP | PADDW |
| 5 | FSUB | JG | FSUBRP | INC | SETNLE | PUSH | MUL | CVTTSS2SI |
| 6 | XCHG | AND | SETNB | SETNLE | JB | CALL | SETNBE | FRNDINT |
| 7 | POP | FIDIV | FILD | SETLE | FMUL | ADD | SETBE | XORPD |
| 8 | OR | FMUL | ROR | MUL | FSTSW | DEC | AND | FDIVRP |
| 9 | FUCOMPP | JL | CALL | FMUL | FADDP | JL | XCHG | ADDPS |
| 10 | JLE | DEC | FISTP | STOS | FSUB | JLE | FSUB | FSUBRP |
| 11 | CMOVS | SUB | FIDIV | FSUB | FISTP | JNB | FISTP | FUCOM |
| 12 | ROR | OR | FNINIT | JB | MUL | AND | JB | PADDD |
| 13 | FIDIV | SETNBE | FSTP | JLE | SHLD | RETN | FSTSW | ORPD |
| 14 | SETBE | FDIV | BSWAP | RCR | FXCH | XOR | SETNLE | PMINUW |
| 15 | JA | FDIVRP | FSTSW | SETZ | SCAS | SUB | JA | COMISS |
| 16 | LEA | FISTP | MUL | NOT | FDIV | TEST | STOS | FIADD |
| 17 | FNINIT | FMULP | FLD | FSTSW | FABS | JZ | FMUL | PUNPCKLBW |
| 18 | CALL | CMC | LODS | FADDP | FILD | JB | SETNB | CMOVBE |
| 19 | AND | NEG | FDIV | JS | FSUBRP | STD | JLE | CMOVB |
| 20 | SETLE | IMUL | CMOVS | JNZ | SETZ | JNS | FSTP | FXAM |
| 21 | FDIVP | LDMXCSR | FSUB | SETNL | FLDZ | OR | SETZ | SUBPS |

[1] PrincipalComponents listed are unranked.

Highlighted in Table 5-16 are any occurrences of the common crypto functions used by ransomware: XOR, ROL, ROR, ROT. Although XOR, ROL, ROR have been selected by 4, 6 and 8 attribute filters respectively as predictive features, only ROR appears more than once in the top 21 ranked predictors and does not feature in the top 20 overall ranked predictors.

The CFSSubset Evaluator can provide feature reduction from 443 to 10 attributes but the ROR attribute can be discarded without any decrease in precision.

To achieve an overall ranking for OpCode predictor importance the top 21 attributes selected by the first 7 evaluation methods in Table 5-17 have been assigned a weight from 21 -> 1, where 21 represents the highest rank, down to 1 for the lowest. The overall ranking is provided in Table 23.

*Table 22*
*Overall ranking for opcode predictor importance*

| Overall Ranking | Opcode | Description [58] | Histogram density % | Histogram density rank /443 |
|---|---|---|---|---|
| 1 | FDIVP | Divide, store result and pop the register stack. | 0.002 | 140 |
| 2 | AND | Logical AND | 1.201 | 15 |
| 3 | SETLE | Set byte if less or equal | 0.014 | 69 |
| 4 | XCHG | Exchange Register/Memory with Register | 0.038 | 54 |
| 5 | SETNBE | Set byte if not below or equal | 0.001 | 145 |
| 6 | SETNLE | Set byte if not less or equal | 0.058 | 45 |
| 7 | JB | Jump short if below | 0.243 | 28 |
| 8 | FILD | Load Integer | 0.035 | 56 |
| 9 | JLE | Jump short if less or equal | 0.244 | 27 |
| 10 | POP | Pop a Value from the Stack | 4.01 | 5 |
| 11 | CALL | Call Procedure | 7.881 | 3 |
| 12 | FSUB | Subtract | 0.025 | 62 |
| 13 | FMUL | Multiply | 0.055 | 46 |
| 14 | MUL | Unsigned Multiply | 0.024 | 63 |
| 15 | SETBE | Set byte if below or equal | 0.001 | 184 |
| 16 | FISTP | Store Integer | 0.008 | 88 |
| 17 | FSUBRP | Reverse Subtract | 0.002 | 138 |
| 18 | INC | Increment by 1 | 0.857 | 17 |
| 19 | FIDIV | Divide | 0.006 | 162 |
| 20 | FSTSW | Store x87 FPU Status Word | 0.008 | 86 |
| 21 | JA | Jump short if above | 0.133 | 36 |

The 2 opcodes with the highest density, MOV (32.45%) and PUSH (13.49%) do not have good predictive importance due to their prevalence in both ransomware and benign samples. Table 4-16 illustrates that some of the more infrequent Opcodes such as SETBE and FIDIV are better indicators of ransomware. This partly agrees with Bilar's conclusion that less frequent opcodes make better indicators of malware than the most frequent opcodes. [23]

## 5. Conclusion

This research demonstrated that the analysis of CPU instructions (opcodes) can be used to differentiate between crypto-ransomware and goodware with high precision.
As per the results presented in this chapter a high precision (>95%) rate has been achieved for the 2-class model using all 4 kernel options, the 6-class model with the PUK kernel, and 6 out of 8 feature reduction models (with the remaining 2 at 94.2%)

It has also been demonstrated that the PUK kernel is the simplest kernel to use for the SMO classifier as it is flexible and self-optimising, achieving 100% precision with no tuning required. The linear, polynomial and Gaussian kernels can also achieve 100% precision when optimised. For all other models (6-class and feature reduction) the PUK kernel achieved the highest precision using the default settings.

By employing the chosen methodology of static analysis, opcode extraction and density histogram representation, a Support Vector Machine can be trained to differentiate between 2 classes (crypto-ransomware and goodware) with 100% precision when using all kernel selections, and between 6 classes (5 crypto-ransomware families and 1 goodware) with 96.5% precision (96.7% accuracy) using the PUK kernel. Both levels of precision exceeded the 95% target set in the objectives. Moreover, when differentiating between ransomware and goodware, feature reduction from 443 extracted opcodes down to 180 opcodes can be achieved using the CorrelationAttributeEval filter with no loss of precision. Feature reduction from 443 to 10 opcodes can be achieved using the CFSSubsetEval filter, but with a lower precision of 94.2%.

There is scope to extend and develop this research. The dataset can be extended to include other crypto-ransomware families such as WannaCry or similar advanced "ransomworms" [59] ; a reduced, optimum feature extraction process can be developed by extracting only the groups of opcodes identified by the attribute selection evaluators; dynamic runtime extraction of opcodes with time-based features can be applied to demonstrate real-time application as a crypto-ransomware threat detection method.

## Acknowledgements

The authors would like to Virus Total for providing access to their Intelligence platform to assist with the dataset creation, and Ransomware Tracker for being an invaluable resource for current ransomware threat detection.